\documentclass[a4paper]{article}

\usepackage{amsmath,amssymb,amsthm,a4wide}
\usepackage[ruled]{algorithm}
\usepackage[noend]{algpseudocode}
\usepackage{enumerate,hyperref,verbatim}
\usepackage{lmodern}
\usepackage[T1]{fontenc}
\DeclareMathOperator{\eval}{word}
\DeclareMathOperator{\first}{first}

\newtheorem{theorem}{Theorem}
\newtheorem{lemma}{Lemma}
\theoremstyle{remark}
\newtheorem{remark}{Remark}

\floatname{algorithm}{\selectfont\sffamily\bfseries Algorithm}

\newcommand{\AC}{{AC}}
\newcommand{\PC}{{PC}}

\newcommand{\polyn}{{\sf poly}\ensuremath{(n)}}
\newcommand{\npoly}{{\sf npolytime}}
\newcommand{\poly}{{\sf polytime}}

\newcommand{\NPclass}{{\sf NP}}
\newcommand{\Pclass}{{\sf P}}
\newcommand{\PSPACE}{{\sf PSPACE}}

\providecommand{\Ocomp}{\mathcal{O}}

\newcommand{\seepage}[1]{\marginpar{(see page~\pageref{#1})}}

\newcommand{\twodots}{\mathinner{\ldotp\ldotp}}

\newcommand{\SLPref}[1]{(\hyperref[#1]{SLP \ref{#1}})}
\newcommand{\Autref}[1]{(\hyperref[#1]{Aut \ref{#1}})}

\newcommand{\SLPrefall}{\SLPref{no new nonterminals}--\SLPref{terminating symbols}}
\newcommand{\Autrefall}{\Autref{letter or nonterminal}--\Autref{starting and accepting state}}

\newcommand{\refall}{\SLPref{no new nonterminals}--\Autref{starting and accepting state}}

\newtheorem{clm}{Claim}

\title{Compressed Membership for NFA (DFA) with
Compressed Labels is in NP (P)
}

\author{Artur Je\.z \\
University of Wroc{\l}aw \\
ul.\ Joliot-Curie~15, 50-383 Wroc{\l}aw, Poland\\
\texttt{aje@cs.uni.wroc.pl}}

\begin{document}

\maketitle

\begin{abstract}
In this paper, a compressed membership problem 
for finite automata, both deterministic and non-deterministic,
with compressed transition labels is studied.
The compression is represented by straight-line programs (SLPs), i.e.
context-free grammars generating exactly one string.
A~novel technique of dealing with SLPs is introduced:
the SLPs are recompressed, so that substrings of the input text
are encoded in SLPs labelling the transitions of the NFA (DFA)
in the same way, as in the SLP representing the input text.
To this end, the SLPs are locally decompressed 
and then recompressed in a uniform way.
Furthermore, such recompression induces only small changes 
in the automaton, in particular, the size of the automaton remains polynomial.

Using this technique it is shown that the compressed membership for NFA
with compressed labels is in NP, 
thus confirming the conjecture of Plandowski and Rytter~\cite{SLPmatchingNFA}
and extending the partial result of Lohrey and Mathissen~\cite{LohreySLP};
as it is already known, that this problem is \NPclass-hard,
we settle its exact computational complexity.
Moreover, the same technique applied to the compressed membership
for DFA with compressed labels yields that this problem is in P;
for this problem, only trivial upper-bound \PSPACE{} was known.
\end{abstract}

\section{Introduction}
\label{sec:intro}
\subsection{Compression and Straight-Line Programms}

Due to ever-increasing amount of data, compression methods
are widely applied in order to decrease the data's size.
Still, the stored data sometimes need to be processed and 
decompressing it on each occasion is wasteful,
Thus there is a large demand for algorithms working directly on the compressed data.
Such task is not as desperate, as it may seem:
it is a popular outlook, that compression basically extracts the hidden
structure of the text and if the compression rate is high,
the text must have a lot of internal structure.
And it is natural to assume, that such a structure
will help devising methods dealing directly
with the compressed representation of the data.
Indeed, efficient algorithms for fundamental text operations
(pattern matching, checking equality, etc.) are known
for various practically used compression methods (LZ, LZW, etc.)~\cite{GawryLZW,GawryLZ,RytterSWAT}.

When devising algorithms for compressed data,
quite early one needs to focus on the exact compression method,
to which the algorithm is applied.
The most practical, and challenging,
choice is one of the widely used standards, like LZW or LZ.
However, a different approach was also proposed: for some applications and for
most of theory-oriented considerations it would be useful to \emph{model} one of the
practical compression standard by a more mathematically well-founded method.
This idea, among other, lay at the foundations of the notion of
\emph{Straight-Line Programms} (SLP), whose instance can be simply seen
as context-free grammars generating exactly one string.


SLPs are the most popular theoretical model of compression.
This is on one hand motivated by a simple, `clean' and appealing definition,
on the other hand, they model the LZ compression standard:
each LZ compressed text can be converted into an equivalent SLP with only logarithmic
overhead (and in polynomial time)
while each SLP can be converted to an equivalent SLP with just a constant overhead
(an in polynomial time).

The approach of modelling compression by SLP in order to develop
efficient algorithms turned out to be fruitful.
Algorithmic problems for SLP-compressed input strings were
considered and successfully solved~\cite{LifshitsClassicCompressed,LifshitsLohrey,PlandowskiSLPequivalence}.
The recent state-of-the-art efficient algorithms for pattern matching
in LZ and LZW compressed text essentially use the reformulation of
LZW and LZ methods in terms of SLPs~\cite{GawryLZW,GawryLZ}.
SLPs found their usage also in programme verification~\cite{GenestMuscholl,RytterLasota}.
Attempts were made in order to make out of the box tools for
texts represented by SLPs, e.g., efficient indexing structure for SLPs
was recently developed~\cite{LZindexing}.

Surprisingly, while SLPs were introduced mainly as a model for practical
applications, they turned out to be	useful also in strictly theoretical branches
of computer science, for instance, their usage was important in the famous
proof of Plandowski, that satisfiability of word equations is in \PSPACE~\cite{PlandowskiEquations}.

\subsection{Membership problem}
As it should be already clear, that SLPs \emph{are} used,
both in theoretical, and applied research in computer science.
Hence, tools for them should be developed.
In particular, one should be aware, that whenever working with strings,
these strings may be supplied as respective SLPs.
Hence, all the usual string problems
should be reinvestigated in the compressed setting,
as the classical algorithms may not apply directly, be inefficient or
worse, some of these problems may become computationally difficult.

From language theory point of view,
the crucial questions stated in terms of strings,
is the one of compressed string recognition.
To be more precise, we consider classic membership problems,
i.e. recognition by automata, generation by a grammar etc.,
in which the input is supplied as an SLP.
We refer to such problems as \emph{compressed membership problems}.
These were first studied in the pioneering work of Plandowski and Rytter~\cite{SLPmatchingNFA},
who considered compressed membership problem for various
formalism for defining languages.
Already in this work it was observed, that we should precisely specify,
what part of the input is compressed?
Clearly the input string, but what about the language representation
(i.e. regular expression, automaton, grammar, etc.).
Should it be also compressed or not?
Both variant of the problem are usually considered,
with the following naming convention: when only the input string is compressed,
we use a name \emph{compressed membership}, when also the language representation,
we prepend \emph{fully} to the name.

In years to come, the compressed membership problem was investigated 
for various language classes
\cite{RytterSWAT,CSR2009,LohreySICOMP,LohreyIJFCS,SLPmatchingNFA}.
Compressed word problem for groups
and monoids \cite{LohreySICOMP,LohreyS07,MacDonald10},
which can be seen as a generalisation of membership problem,
was also investigated.

Despite the large attention in the research community,
the exact computational complexity of some problems remain open.
The most notorious of those is the fully compressed membership problem for NFA,
considered already in the work of Plandowski and Rytter~\cite{SLPmatchingNFA}.
Here, the compression of NFA is done by allowing
the transitions by strings, and only by single letters,
and representing these strings as SLPs.
%

It is relatively easy to observe
that the compressed membership problem for the NFA is in P,
however, the status of the fully compressed variant remained open for a long time.
Some partial results were obtained by
Plandowski and Rytter~\cite{SLPmatchingNFA},
who observed that it is in \PSPACE{} and is \NPclass-hard
for the case of unary alphabet, both of these bounds being relatively simple.
Moreover, they showed that this problem is in \NPclass{} for some particular cases,
for instance, in the case of one-letter alphabet.
Further work on the problem was done by Lohrey and Mathissen~\cite{LohreySLP},
who demonstrated that if the strings defined by SLP
have polynomial periods, the problem is in \NPclass, and when all strings are highly aperiodic,
it is in \Pclass.

\subsection{Our results and  techniques}
We establish the computational complexity of fully compressed membership problems
for both NFAs and DFAs.
\begin{theorem}
\label{thm:NFA in NP}
Fully compressed membership problem for NFA is in \NPclass,
for DFA it is in \Pclass.
\end{theorem}

Our approach to the problem is essentially different than the approach
of Plandowski and Rytter~\cite{SLPmatchingNFA} and Lohrey and Mathissen~\cite{LohreySLP}.
The main ideas utilised in the previous approaches focused
on the properties of strings described by SLPs.
We take a completely different approach:
we analyse and change the way strings are described by the SLPs in instance.
That is, we focus on the SLPs, and not on the encoded strings.
Roughly speaking, our algorithm aims at having all the strings in the instance
compressed `in the same way'.
To achieve this goal, we decompress the SLPs.
Since the compressed text can be exponentially long, we do this locally:
we introduce explicit strings into the right-hand sides of the productions.
Then, we recompress these explicit strings uniformly.
Since such pieces of text are compressed in the same way,
we can `forget' about the original substrings of the input
and treat the introduced nonterminals as atomic letters.
The idea is that such recompression
should shorten the text significantly:
roughly one `round' of recompression,
in which every pair of letters that was present at the beginning of the `round'
is compressed,
should shorten the encoded strings by a constant factor.
It remains to implement the recompression (and changes in the NFA)
in \npoly, while keeping the size of $N$ polynomial.

We stress, that the idea of local decompression
and synchronous recompression of SLP is new and promising:
there is hope that it can be applied
to other problem related to SLPs.

\section{Preliminaries}
\label{sec:prelim}
\subsection{Straight line programmes}
Formally, a \emph{Straight line programme} (SLP) is context free grammar $G$
over the alphabet $\Sigma$,
with a language consisting of exactly one string.
Usually it is assumed that $G$ is in a \emph{Chomsky normal form},
i.e. each production is either of the form $X \to YZ$ or $X \to a$.
This assumption in particular implies,
that strings defined by nonterminals have length at most $2^n$;
since our algorithm will replace some substrings by shorter ones, none
string defined by SLPs during the run of algorithm will exceed this length.

We denote the string defined by nonterminal $A$ by $\eval(A)$.
This notion extends to $\eval(\alpha)$ for $\alpha \in (\mathcal X \cup\Sigma)^*$ in the usual way.

\subsection{Input}
The instance of the fully compressed membership problem for NFA
consists of an input string, represented by an SLP,
and an NFA $N$, whose transitions may be labelled by SLPs.

For our purposes it is more convenient to assume,
that all SLPs are given as a single context free grammar $G$
with a set of nonterminals $\mathcal X = \{ X_1, \ldots , X_n \}$,
the input string is defined by $X_n$
and the NFAs transitions are labelled with nonterminals of $G$.
Furthermore, in our proof, it is essential to drop the usual assumption,
that $G$ is in a Chomsky normal form.
However, we still impose some conditions on the productions' right-hand sides.
Thus, we require that the grammar during the run of algorithm
satisfies the following constraints on its form:
\begin{subequations}
\label{form}
\begin{align}
	\label{one production}
	&\text{each nonterminal has exactly one production, which is of the form}\\
\label{eq:rule form}
	& X_i \to u X_j v X_k \; \text{ or } \;
	X_i \to u X_j v \; \text{ or } \;
	X_i \to u, \quad
\text{where $u,v \in \Sigma^*$ and $j,k < i$},\\
\label{trivial epsilon rules}
	&\text{if } \eval(X_i) = \epsilon \text{ then } X_i \text{ is not 
	on the right-hand side of any production}.
\end{align}
\end{subequations}

The strings $u$, $v$ and their substrings appear \emph{explicitly} in a rule,
this notion is introduced to distinguish them from the substrings of $\eval(X_i)$.
Notice, that~\eqref{form} does not exclude the case,
when $X_i \to \epsilon$ and allowing such a possibility streamlines the analysis.

Without loss of generality we may assume that the input string starts
and ends with designated, unique symbols, denoted as $\$$ and $\#$.
These are not essential, however, the first and last letter of $\eval(X_n)$
need to be treated in a somewhat special manner, furthermore,
this applies to their appearances in the NFA as well.
Having special symbols for the first and last letter makes the analysis smoother.

\subsection{Input size, complexity classes}
The size $|G|$ of the representation of grammar $G$ is the sum of length
of the right-hand sides of $G$'s rules.
The size $|N|$ of the representation of NFA $N$
is the sum of number of its states and transitions.
The size $|\Sigma|$ of alphabet $\Sigma$ is simply the number of elements in $\Sigma$.

The input (or, in general, current instance)
size is polynomial in $N$, $G$, $\Sigma$ and $n$,
which denotes the number of nonterminals in $G$.
We point out, that one of the crucial properties
of our algorithm is that $n$ is not modified during the whole run of the algorithm.

By \npoly{} (\poly) we denote the class of algorithms running in non-deterministic (deterministic, respectively)
polynomial time, and by \NPclass{} (\Pclass, respectively)
the corresponding complexity classes of the decision problems.


\subsection{Known results}
We use the following basic result, which states
that the fully compressed membership problem,
when the input string is over a unary alphabet, is in \NPclass{} for NFA 
and in \Pclass{} for DFA.
\begin{lemma}[cf.~{\cite[Theorem~5]{SLPmatchingNFA}}]
\label{lem:unary case is in NP}
The fully compressed membership problem restricted to the input string
over an alphabet $\Sigma = \{ a\}$ is in \NPclass{} for NFA and in \Pclass{} for DFA.
\end{lemma}
The first claim can be easily inferred from the result of
Plandowski and Rytter~\cite[Theorem~5]{SLPmatchingNFA},
who proved that this problem is in NP, when $\Sigma = \{ a\}$,
i.e. also transitions in the NFA are labelled by powers of $a$ only.
The second claim is trivial.

\subsection{Path and labels in NFA}
Since we deal with automata, proofs will be concerned with
(accepting) paths for strings.
We shall consider NFAs, for which transitions are labelled with either letters,
or non-terminals of $G$.
That is, that transition relation $\delta$ satisfies
$\delta \subseteq Q \times (\Sigma \cup \mathcal X) \times Q$.
Consequently, a \emph{path} $\mathcal P$ from state $p_1$ to $p_{k+1}$
is a sequence $\alpha_1\alpha_2\dots\alpha_k$,
where $\alpha_i \in \Sigma \cup \mathcal X$  and $\delta(q_i,\alpha_i,q_{i+1})$.
We write, that $\mathcal P$ \emph{induces} such a list of labels.
The $\eval(\mathcal P)$ \emph{defined} by such a path $\mathcal P$ is
simply $\eval(\alpha_1\dots \alpha_k)$.
We also say that $\mathcal P$ is a path for a string $\eval(\mathcal P)$.
A path is \emph{accepting}, if it ends in an accepting state.
A string $w$ is accepted by $N$ if there is an accepting path from the starting state for $w$.

\section{Basic classifications and outline of the algorithm}
\label{sec:outline}
In this section we present the outline of the algorithm
for fully compressed membership for NFAs.
Its main part consist of recompression, i.e. replacing strings appearing
in $\eval(X_n)$ by shorter ones.
In some cases, such replacing is harder, in other easier.
It should be intuitively clear, that it depends on the position of
letters inside encoded texts:
if $a$ is a first or last letter of some $\eval(X_i)$, then recompressing
strings including $a$ looks difficult, as 
such strings can be split into different nonterminals and recompression
requires heavy modification of $G$, or even rebuilding of the NFA.
On the other hand, if a letter $a$ is only `inside' strings encoded by nonterminals,
its compression is done only `inside' rules of $G$, which seems easy.
Thus, before we state the algorithm, we firstly introduce classification
of letters (and strings) into `easy' and `difficult' to compress.

\subsection{Crossing appearances, types of letters, non-extendible appearances}
We say that a string $w$ has a \emph{crossing appearance in a (string defined by)
nonterminal} $X_i$ with a production $X_i \to uX_jvX_k$,
if $w$ appears in $\eval(X_i)$, but this appearance is not contained in
neither $u$, $v$, $\eval(X_j)$ nor $\eval(X_k)$.
Intuitively, this appearance `crosses' the symbols in
of $u\eval(X_j)v\eval(X_k)$,i.e, at the same time
part of $w$ is in the explicit substring ($u$ or $v$) and part
is in the compressed strings ($\eval(X_j)$ or $\eval(X_k)$).
This notion is similarly defined for nonterminals
with productions of the form $X_i \to u X_j v$,
productions of the form $X_i \to u$ clearly do not have crossing appearances.

A string $w$ has a \emph{crossing appearance in the NFA} $N$,
if there is a path in $N$ inducing list of labels $\alpha_1\alpha_2$,
where $\alpha_1, \alpha_2 \in \{X_1, \ldots, X_n \} \cup \Sigma$
with at least one of $\alpha$, $\alpha_2$ being a nonterminal,
such that $w$ appears in a $\eval(\alpha\alpha_2)$,
but this appearance is not contained in the $\eval(\alpha_1)$,
nor in $\eval(\alpha_2)$.
The intuition is similar as in the case of crossing appearance in a rule:
it is possible that a string $w$ is split between
two transitions' labels.
Still, there is nothing difficult in consecutive letter transitions,
thus we treat such a case as a simple one.

We say that a pair of different letters $ab$ is a \emph{crossing pair},
if $ab$ has a crossing appearance of any kind.
Otherwise, such a pair is \emph{non-crossing}.

We say that a letter $a \in \Sigma$ is \emph{right-outer} (\emph{left-outer}),
if there is a nonterminal $X_i$, such that $a$ is the leftmost (rightmost, respectively) symbol in
$\eval(X_i)$.
A letter is \emph{outer}, if it is left-outer or right-outer.
Otherwise, the letter is \emph{inner}.
Notice, that if a pair $ab$ is crossing,
then $a$ is left-outer or $b$ is right-outer.
The outer letters and crossing pairs correspond to the intuitive notion
of `hard' to compress.

The following lemma shows, that while $G$ may encode long strings,
they have relatively few different short substrings and few outer letters.

\begin{lemma}
\label{lem:different crossing}
There are at most $2n$ different outer letters
and at most $|G|+3n$ different pairs of letters appearing in $\eval(X_n)$, \ldots, $\eval(X_n)$.

The set of outer letters, the set of crossing pairs and the set of non-crossing pairs
appearing in $\eval(X_n)$, \ldots, $\eval(X_n)$ can be computed in \poly{}.
\end{lemma}

The notions of (non-) crossing pairs do not apply to $aa$,
still, an analog can be defined:
for a letter $a \in \Sigma$ we say that $a^\ell$ is a $a$'s \emph{non-extendible appearance} of length $\ell$,
if it appears in some string defined by some nonterminal and it is surrounded by letters other than $a$,
formally, if there exist two letters $x ,y \in \Sigma$, where $x \neq a \neq y$ and a nonterminal $X_i$,
such that $xa^\ell y$ is a substring of $\eval(X_i)$.
Similarly to crossing pairs, it can be shown that there are not too many different
non-extendible appearances of $a$.

\begin{lemma}
\label{polynomial non extendible appearances}
For an inner letter $a$ and a grammar $G$
there are at most $|G|$ different lengths of $a$'s non-extendible appearances
in $\eval(X_1)$, \ldots, $\eval(X_n)$.
The set of these lengths can be calculated in \poly.
\end{lemma}

\subsection{Outline of the algorithm}
Our algorithm consists of two main operations performed on strings encoded
by $G$
\begin{description}
	\item[appearance compression of $a$] 
	For each $a^\ell$ that has a non-extendible appearance in $\eval(X_n)$,
	replace all $a^\ell$s in $\eval(X_1)$, \ldots, $\eval(X_n)$
	by a fresh letter $a_\ell$. Modify $N$ accordingly.
	\item[pair compression of $ab$]
	For two \emph{different} letters $ab$ replace each of $ab$
	in $\eval(X_1)$, \ldots, $\eval(X_n)$ by a fresh letter $c$.
	Modify $N$ accordingly.
\end{description}
We denote the string obtained from $w$ by compression of appearances of $a$ by $\AC_a(w)$,
and the string obtained by compression of a pair $ab$ into $c$ by $\PC_{ab \to c}(w)$.

We adopt the following notational convention throughout rest of the paper:
whenever we refer to a letter $a_\ell$, it means that the last appearance compression
was done for $a$ and $a_\ell$ is the letter that replaced $a^\ell$.

The main idea behind the algorithm is that appearance compression and pair compression
shorten the encoded texts significantly.
The challenging part of the algorithm is the modification of NFA $N$ `accordingly' to the changes of SLPs.
The general schema is given by Algorithm~\ref{alg:main}.
\begin{algorithm}[H]
  \caption{Outline of the main algorithm}
  \label{alg:main}
  \begin{algorithmic}[1]
  \While{$|\eval(X_n)>n|$} \label{alg:mainloop}
    \While{something changed}
    	\For{$a$: inner letter}
    			\State \label{ac noncrossing} compress appearances of $a$, modify $N$ accordingly
    	\EndFor
    	\For{non-crossing pair $a b$ in $\eval(X_n)$, $a, \notin \{ \$ , \# \}$}
    		\State \label{pc noncrossing} compress $ab$, modify $N$ accordingly
    	\EndFor
		\EndWhile

		\State $L \gets $ list of outer letters, except $\$$ and $\#$ 		\label{cr list}
		\For{$a \in L$}
			\State \label{ac crossing} compress appearances of $a$, modify $N$ accordingly
			\For{each $a_\ell b$ in $\eval(X_n)$}
				\State \label{pc crossing} compress $a_\ell b$, modify $N$ accordingly
			\EndFor
		\EndFor
		\EndWhile
	\State Decompress $X_n$ and solve the problem naively.
	 \end{algorithmic}
\end{algorithm}

There are two important remarks to be made:
\begin{itemize}
	\item there is no explicit non-deterministic operation in the code,
	however, it appears implicitly in the term `modify the NFA accordingly'
	in lines~\ref{ac noncrossing} and~\ref{ac crossing}.
	Roughly, one needs to solve
	fully compressed membership for string $a^\ell$ for that, and this is
	known to be \NPclass-hard.	
	\item the compression (both of pairs and appearances) is never applied to $\$$, nor to  $\#$.
	The markers were introduced so that we do not bother with strange behaviour
	where first or last letter is compressed, and so we do not touch the markers.
\end{itemize}

It should be more or less obvious, that the compressions performed
by Algorithm~\ref{alg:main} shortens $\eval(X_n)$.
This is formally stated as follows.
\begin{lemma}
\label{lem:logM iterations}
There are 
$\Ocomp(n)$ executions of the loop in line~\ref{alg:mainloop} of Algorithm~\ref{alg:main}.
\end{lemma}

\begin{remark}
Notice, that pair compression $\PC_{ab \to b}$ is in fact introducing a new
nonterminal with a production $c \to ab$, similarly $\AC_a$.
Hence, Algorithm~\ref{alg:main} creates new SLPs,
encoding strings from the instance.
However, these new nonterminals are never expanded,
they are always treated as individual symbols.
Thus it is better to think of them as letters.
In particular, the analysis of running time of Algorithm~\ref{alg:main}
relies on the fact, that no new nonterminals are introduced by Algorithm~\ref{alg:main}.
\end{remark}

\section{Details}
\label{sec:details}
In this section we describe in detail how to implement the appearance compression
and pair compression.
In particular, we are going to formulate the connections between NFA and SLPs
preserved during Algorithm~\ref{alg:main} and demonstrate how to modify the NFA accordingly.

\subsection{Invariants}
The invariants below describe the connection between the grammar kept by Algorithm~\ref{alg:main}
and the input one.

\begin{enumerate}[SLP 1]
	\item \label{no new nonterminals}
	The set of used nonterminals is a subset of $\mathcal X = \{X_1, \ldots , X_n\}$
	and the productions are of the form described in~\eqref{form}.
	\item \label{no new productions}
	For every production $X_i \to \alpha_i$, the original
	instance contained a production $X_i \to \alpha_i'$, where
	the nonterminals appearing in $\alpha$ appear (in the same order) in $\alpha'$.
	\item \label{terminating symbols}
	The nonterminal $X_n$ has a production $X_n \to \$ u X_{n-1} v \#$,
	where $u,v \in (\Sigma \setminus \{ \$, \# \})^*$;
	$\$ $, $\#$ are not used in other productions.
\end{enumerate}
The following invariants represent the constraints on the NFA.
\begin{enumerate}[{Aut} 1]
	\item \label{letter or nonterminal} every transition of $N$
	is labelled by a single letter of $\Sigma$ (\emph{letter transition})
	or by a nonterminal (\emph{nonterminal transition}),
	each nonterminal labels at most one transition.
	No transition is labelled with $X_n$.
	\item \label{starting and accepting state}
	there is a unique starting state that has a unique outgoing transition, by letter $\$$,
	and no incoming transitions;
	there is no other transition by $\$$.
	Similarly, there is a unique accepting state that has a unique incoming transition, by letter $\#$,
	it does not have any outgoing transitions;
	there is no other transition by $\#$ in $N$.
\end{enumerate}
Algorithm~\ref{alg:main} will preserve \refall,
and we shall always assume, that the input of the subroutines
satisfies \refall.

We assume that the input instance satisfies \refall,
moreover, that that the input grammar is in the Chomsky normal form.
It is routine to transform (in \poly) the input instances not satisfying these
conditions into equivalents instance that satisfy them.

\subsection{Compression of non-crossing pairs and inner letters}
The compression of non-crossing pairs and appearance compression for inner letters
is intuitively easy: whenever these appear
in strings encoded by $G$ or on paths in $N$,
they cannot be split between nonterminals or between transitions.
Thus, it should be enough to replace their explicit appearances in the grammar and in the NFA.
This is formalised and shown in this subsection

\subsubsection{Compression of non-crossing pairs}
We first demonstrate, how to perform the pair compression for non-crossing pairs.
Consider a non-crossing pair $ab$.
Since it is non-crossing, it can only appear in the
the explicit strings in the rules of $G$.
Hence, compressing $ab$ into a fresh letter $c$
consists simply of replacing each explicit $ab$ by $c$
in right-hand side of every production.
Still, $ab$ can appear on a path in $N$.
But since $ab$ is non-crossing, this can be either wholly inside
a nonterminal transition (and so compression was already taken care of),
or on two consecutive letter transitions.
This is also easy to handle: whenever there is a path from $p$ to $q$ by a string
$ab$, we introduce a new letter transition by $c$ from $p$ to $q$.
This description is formalised in Algorithm~\ref{pc code}.

To distinguish between the input and output $G$ and $N$,
we utilise the following convention:
`unprimed' names refer to the input
(like $G$, $X_i$, $N$), while `primed' symbols refer to the output (like $G$, $X_i'$, $N'$).
This convention is used for lemmata concerning algorithms through the paper.
\begin{algorithm}[H]
  \caption{Pair compression for a non-crossing pair $ab$ \label{pc code}}
  \begin{algorithmic}[1]
  \For{each production $X_i \to \alpha$} \Comment{Pair compression}
  \label{pc new letter}
  	\State replace each explicit $ab$ in $\alpha$ by $c$
  \EndFor
  \For{states $p$, $q$}	\Comment{Appropriate modifications of $N$}
  	\If{there is a path for $ab$ from $p$ to $q$}
  		\State put a transition $\delta_{N}(p,c,q)$
  	\EndIf
  \EndFor
 \end{algorithmic}
\end{algorithm}

We show, that Algorithm~\ref{pc code} indeed realises
the pair compression for a non-crossing pair.
\begin{lemma}
\label{lem:pc noncrossing}
Algorithm~\ref{pc code} runs in \poly{} and preserves~\refall.
When applied to a non-crossing pair of letters $ab$, where $a,b \notin \{ \$ , \#\}$,
it implements the pair compression, i.e. $\eval(X_i') = \PC_{ab \to c}(\eval(X_i))$,
for each $X_i$.

$N'$ recognises $\eval(X_n')$ if and only if $N$ recognises $\eval(X_n)$.
If $N$ is a DFA, so is $N'$.
\end{lemma}

\subsubsection{Inner letter appearance compression}
We can apply the same approach, as the one used for non-crossing pairs compression,
to the inner letters appearance compression.
However, in this case, the modification of $N$ uses non-determinism.

Since $a$ is an inner letter, it cannot
appear as the last or first letter of any nonterminal, and so every
non-extendible appearance of $a$ in $\eval(X_1)$, \ldots, $\eval(X_n)$
is an explicit substring in the right-hand sides of $G$'s rules;
so we simply replace explicit $a^\ell$ by a fresh letter $a_\ell$ in each right-hand
side of $G$'s rules.
Before considering the NFA, notice that as $a$ is an inner letter,
$a^\ell$ cannot have a crossing appearance in $N$,
and no nonterminal defines $a^\ell$.
Hence, when $a^\ell$ is a substring of a string defined by a path in $N$,
then $a^\ell$ appears wholly inside a nonterminal transition,
and this was already taken care of when considering $G$,
or $a^\ell$ labels a path using letter transitions only.
So it is enough to check,
whether there is a path for $a^\ell$ from $p$ to $q$ using
only letter transitions.
%
\begin{algorithm}[H]
  \caption{Appearance compression for an inner letter $a$ \label{ac code}}
  \begin{algorithmic}[1]
  \State \label{establish lengths} establish the lengths $\ell_1, \ldots, \ell_k$ of $a$'s
  non-extendible appearance
  \For{each $a^{\ell_m}$} \Comment{Appearance compression}
		\For{each production $X_i \to \alpha$}
  		\State \label{replace non-extendible}
		replace every explicit non-extendible $a^{\ell_m}$ in in $\alpha$ by $a_{\ell_m}$
  	\EndFor  
  	\For{states $p$, $q$ in $N$}
  	\Comment{Appropriate modifications of $N$}
  			\If{$\delta_N(p,a^{\ell_m},q)$} \label{transition guess}
  			\Comment{Verify non-deterministically, %
  			see Lemma~\ref{lem:unary case is in NP}}
  				\State put a transition $\delta_{N}(p,a_{\ell_m},q)$
  			\EndIf
  	\EndFor
  \EndFor
  \end{algorithmic}
\end{algorithm}
%
%
%
%


\begin{lemma}
\label{lem:ac inner}
Suppose that Algorithm~\ref{ac code} is applied for inner letter $a \notin \{\$,\#\}$.
Then Algorithm~\ref{ac code} properly implements non-extendible appearance compression,
i.e. $\eval(X_i') = \AC_a(\eval(X_i))$ for each $X_i$ and
	preserves~\refall. 

The operations in line~\ref{transition guess} of Algorithm~\ref{ac code}
can be performed in \npoly{}, other operations can be performed in \poly.

Each of the new letters $a_\ell$ is inner.
$N$ recognises $\eval(X_n)$ if and only if
$N'$ recognises $\eval(X_n')$ for some non-deterministic choices.
If $N$ is DFA, so is $N'$.
\end{lemma}

\subsection{Compression of outer letters and crossing pairs}
Now, we turn our attention to the compression of outer letters and crossing pairs.
%
%
The outline is as follows:
we fix an outer letter $a$ and modify the instance,
so that $a$ becomes inner.
Then, Algorithm~\ref{ac code} is applied to $a$.
Next, we want to compress each pair of the form $a_\ell b$.
Such a pair can be crossing, as $b$ can be a left-outer letter.
Thus, we modify the instance again,
so that none of $a_\ell b$ is  a crossing pair
so that it can be compressed using Algorithm~\ref{pc code}.

\subsubsection{Transforming an inner letter to an outer letter}
The reason, why $a$ is an outer letter, is that it is the first
or the last symbol in some $\eval(X_i)$.
To make it an inner letter, it is enough to remove each nonterminal's
$a$-prefix and $a$-suffix. 
To be more precise: fix $i$ and let $\eval(X_i) = a^{\ell_i} u a^{r_i}$,
where $u$ does not start nor end with $a$.
Then our goal is to modify $G$ so that $\eval(X_i') = u$.
(If $\eval(X_i)$ is a power of $a$, we simply give $u' = \epsilon$ and $r_i=0$.)
This can be done in a bottom-up fashion, starting from $X_1$:
it is enough to calculate and memorise the lengths of the $a$-prefixes and $a$-suffixes
for consecutive nonterminals,
see the loop in lines~\ref{replace pop code begins}--\ref{prefix}
of Algorithm~\ref{removing outer letters}. 
%
Then we need to modify the NFA accordingly:
it is enough to replace the transition labelled with $X_i$
by path consisting of three transitions,
labelled with $a^{\ell_i}$, $X_i'$ and $a^{r_i}$.

%
\begin{algorithm}[H]
  \caption{Changing an inner letter $a$ to an outer letter. \label{removing outer letters}}
  \begin{algorithmic}[1]
  \For{$i=1 \twodots n$} \label{replace pop code begins} \Comment{Removing $a$-prefix and suffix}
  	\State let the production for $X_i$ be $X_i \to \alpha_i$ 
  	\State \label{replace pop code}
  	replace any nonterminal $X_j$ in $\alpha_i$ by $a^{\ell_j}X_ja^{r_j}$, remove those defining $\epsilon$
  	\State \label{prefix}
  	calculate the explicit $a$-prefix $a^{\ell_i}$ and $a$-suffix $a^{r_i}$ of $\alpha_i$ and remove them
	\If{there is a nonterminal transition $\delta_N(p,X_i,q)$ in $N$}
	\label{moving pop code begins}
	\Comment{modification of NFA}
				\State
				\label{new states outer letters}
				create new states $p_1$, $q_1$ in $N$, remove transition $\delta_{N}(p,X_i,q)$
				\State
				\label{new transition outer letters}
				set transitions: $\delta_{N}(p,a^{\ell_i},p_1)$,
				$\delta_{N}(p_1,X_i,q_1)$, $\delta_{N}(q_1,a^{r_i},q)$
	\EndIf
  \EndFor
  \end{algorithmic}
\end{algorithm}

The removed $a$-prefixes and $a$-suffixes can be exponentially long,
and so we store them in the rules in a succinct way, i.e. $a^\ell$ is represented as $(a,\ell)$;
the size of representation of $\ell$ is $\Ocomp(\log \ell)$,
that is, linear in $n$.
We say, that such a grammar is in an $a$-\emph{succinct} form.
The situation is similar for the NFA,
as it might have transitions labelled with $a^\ell$, which are stored in succinct way as well.
We say that $N$ satisfies $a$-\emph{relaxed}~\Autref{letter or nonterminal},
if its transitions are labelled by nonterminals,
a single letter or by $a^\ell$,
where $\ell \leq 2^n$.

\begin{lemma}
\label{lem:ai are not crossing}
Algorithm~\ref{removing outer letters} applied to letter $a \notin \{\$, \#\}$
runs in \poly{} time and preserves~\refall,
except that it $a$-relaxes \Autref{letter or nonterminal}.
$G'$ is in the $a$-succinct form.

Let $\eval(X_i) = a^{\ell_i}u_ia^{r_i}$, where $u_i$ does not begin, nor end with $a$.
Then (after running Algorithm~\ref{removing outer letters}), $\eval(X_i') = u_i$.
In particular, after running Algorithm~\ref{removing outer letters}
the letter $a$ is inner.

$N$ accepts $\eval(X_n)$ if and only if $N'$ accepts $\eval(X_n')$.
If $N$ is a DFA, so is $N'$.
\end{lemma}

Since $a$ is no longer an outer letter, we may compress its non-extendible appearances
using Algorithm~\ref{ac code}.
Some small twitches are needed to accommodate the $a$-succinct form of $G$
and the fact that $N$ is $a$-relaxed,
though basically we just perform appearances compression for inner $a$,
as described by Algorithm~\ref{ac code}:
the non-trivial part of Algorithm~\ref{ac code} was the application
of Lemma~\ref{lem:unary case is in NP},
which works for such large powers of $a$ in \npoly.
Other actions of Algorithm~\ref{ac code} generalise in a simple way.

\begin{lemma}
\label{lem:ac extension}
Algorithm~\ref{ac code} can be extended, so that it applies to
instances satisfying \SLPrefall{} with $G$ in the $a$-succinct form
and $a$-relaxed-\Autrefall.
Lemma~\ref{lem:ac inner} applies to such an extension.
The output satisfies~\refall.
\end{lemma}

\subsubsection{Crossing pair compression}
By Lemma~\ref{lem:ac extension} all letters $a_\ell$ are inner.
However, a pair of the form $a_\ell b$ can still be crossing and this can happen only
when $b$ is a left-outer letter,
so we would like to make such $b$ not a left-outer letter.
To do so, we `pop' one letter from the beginning of each nonterminal
(that is, all left outer letters), i.e.
we modify the grammar so that $\eval(X_i) = \first(X_i)\eval(X_i')$,
where $\first(X_i)$ denotes the first letter of $\eval(X_i)$.
Clearly, after such operations there are some (perhaps other)
left-outer letters in $G$.
Still, we show, that none $a_\ell b$ is crossing.

Popping letters is performed similarly
to the removal of the $a$-prefix, i.e. in a bottom-up fashion, starting from $X_1$:
in a rule $X_i \to \alpha$ it is enough to replace each nonterminal $X_j$
in $\alpha$ by $\first(X_j) X_j'$, then store the first letter from $\alpha$
in $\first(X_i)$ and remove it from $\alpha$.
It is easy to modify the NFA $N$ accordingly: when there is a transition
$\delta_N(p,X_i,q)$, we change it into a chain of two transitions:
$\delta_{N'}(p,\first(X_i),p_1)$ and $\delta_{N'}(p_1,X_i',q)$.
This operation is not performed on $X_n$,
as the letter $\$$ is not going to be compressed anyway.
There is a little detail to take care of:
if $|\eval(X_i)|=1$ then popping a letter from $X_i$ creates $X_i'$, which defines $\epsilon$.
Then $X_i'$ should be removed from the right-hand sides of the rules
and in the NFA $N$ we simply replace the transition by $X_i$ by a transition by $\eval(X_i)$.
This description is formalised in Algorithm~\ref{pop code}.

\begin{algorithm}[H]
  \caption{Popping letters \label{pop code}}
  \begin{algorithmic}[1]
  \For{$i\gets 1 \twodots n-1$}  \Comment{Popping letters}
  	\State let $X_i \to \alpha$ 
	\State \label{pop code new letter} replace each $X_j$ in $\alpha$ by $\first(X_j)X_j$,
	remove those defining $\epsilon$
  	\State \label{first letter}
  	set $\first(X_i) \gets $ the first letter of $\eval(X_i)$
  	\If{$i<n$}  	
  		\State remove the first letter from $\alpha$
  	\EndIf
	\If{there is a transition $\delta_N(p,X_i,q)$ in $N$} \Comment{NFA modification}
		\State remove transition $\delta_{N}(p,X_i,q)$
		\If{$|\eval(X_i)| > 1 $}
			\State
			\label{new states pop}
			create new state $p_1$ in $N$,
			set transitions: $\delta_{N}(p,\first(X_i),p_1)$, $\delta_{N}(p_1,X_i,q)$
		\Else
			\State set transition $\delta_{N}(p,\first(X_i),q)$
		\EndIf
	\EndIf
  \EndFor
  \end{algorithmic}
\end{algorithm}

\begin{lemma}
\label{lem:pop}
Algorithm~\ref{pop code} runs in time \poly{}
and preserves~\refall.
Let $\eval(X_i)=bu$, where $b \in \Sigma$,
then $\eval(X_i')=u$ for $i<n$ and $\eval(X_n') = \eval(X_n)$.
After running Algorithm~\ref{pop code},
pairs of the form $a_\ell b$ appearing in $\eval(X_n)$ are non-crossing.

$N'$ accepts $\eval(X_n')$ if and only if $N$ accepts $\eval(X_n)$.
If $N$ is deterministic, so is $N'$.

\end{lemma}

Now, it is enough to apply the pair compression
for non-crossing pairs to each pair of the form $a_\ell b$.
For convenience, we write the whole procedure for pair compression
for crossing pairs in Algorithm~\ref{pc code crossing}.

\begin{algorithm}[H]
  \caption{Pair compression for crossing pairs $a_\ell b$ \label{pc code crossing}}
  \begin{algorithmic}[1]
  \State pop the first letter from each nonterminal (run Algorithm~\ref{pop code})
  \For{each $a_\ell$}
  	\For{each $b$ such that $a_\ell b$ appears in $\eval(X_n)$}
  		\State run Algorithm~\ref{pc code} for a pair $a_\ell b$
  	\EndFor
  \EndFor
  \end{algorithmic}
\end{algorithm}

\begin{lemma}
\label{lem:pc crossing}
Algorithm~\ref{pc code crossing} runs in \poly{} and preserves~\refall.
It implements pair compression for $ab$, in the sense that
$\eval(X_i') = \PC_{ab \to c}(\eval(X_i))$ for each $X_i$.

$N'$ accepts $\eval(X_n')$ if and only if $N$ accepts $\eval(X_n)$.
If $N$ is deterministic, so is $N'$.
\end{lemma}

\subsection{Running time}

Since the running time of each algorithm is \npoly,
it is enough to show that the size of $\Sigma$, $G$ and $N$ are always polynomial in $n$
(recall, that $n$ is unchanged throughout Algorithm~\ref{alg:main}).

\begin{lemma}
\label{lem:running time}
During Algorithm~\ref{alg:main}, the sizes of $\Sigma$, $G$, $N$ are polynomial in $n$.
\end{lemma}

Using Lemmas~\ref{lem:pc noncrossing}--\ref{lem:running time}
it is now possible to conclude that Algorithm~\ref{alg:main}
correctly solves the fully compressed membership problem for NFA,
in nondeterministic polynomial (in $n$) time.
The only source of non-determinism is the one in Lemma~\ref{lem:unary case is in NP},
and so for DFA the corresponding problem can be solved deterministically.

\subsection*{Acknowledgements}
I would like to thank Pawe\l{} Gawrychowski for introducing me to the topic
and for pointing out the relevant literature.

\appendix

\clearpage

\begin{center}
{\huge Appendix}
\end{center}

\section{Additional material for Section~\ref{sec:intro}}
\begin{proof}[proof of Theorem~\ref{thm:NFA in NP}]
\seepage{thm:NFA in NP}
The proof follows by showing that Algorithm~\ref{alg:main}
properly verifies, whether $\eval(X_n)$ is accepted by $N$
and that Algorithm~\ref{alg:main} runs in \npoly.

Let us first show correctness of Algorithm~\ref{alg:main}.
All subroutines of Algorithm~\ref{alg:main} (non-deterministically)
modify the instance, changing $G$, $N$ and $X_n$ into
$G'$, $N'$ and $X_n'$
(notice, that the output depends on the non-deterministic choices).
Let $N^{(i)}$, $G^{(i)}$, $X_n^{(i)}$ for $i=1$, \ldots , $k$
be the consecutive obtained instances,
with $i=1$ representing the input instance.
Then $N^{(i)}$ accepts $\eval(X_n)^{(i)}$
if and only if for some non-deterministic choices
the resulting $N^{(i+1)}$ accepts $\eval(X_n)^{(i+1)}$.
This is shown in Lemmata~\ref{lem:pc noncrossing}, \ref{lem:ac inner},
\ref{lem:ai are not crossing}, \ref{lem:ac extension}, \ref{lem:pop},
\ref{lem:pc crossing} (if some of the procedures are deterministic,
then the output does not depend on any choices).
So, if $N^{(1)}$ does not accept $\eval(X_1^{(1)})$ also
$N^{(k)}$ does not accept $\eval(X_1^{(k)})$.
On the other hand, if
$N^{(1)}$ accepts $\eval(X_n^{(1)})$, then there exits a sequence
of instances (representing proper non-deterministic guesses), such that for each $i$
$N^{(i)}$ accepts $\eval(X_n^{(i)})$.
In particular, $N^{(k)}$ does accept $\eval(X_n^{(k)})$ and
as $|\eval(X_1^{(k)})| < n$, $\eval(X_1^{(k)})$ can be decompressed
and acceptance by $N^{(k)}$ can be checked naively in \poly.

Now, we should show that the running time is in fact (non-deterministic) polynomial.
Lemmata~\ref{lem:pc noncrossing}, \ref{lem:ac inner},
\ref{lem:ai are not crossing}, \ref{lem:ac extension}, \ref{lem:pop},
\ref{lem:pc crossing} claim, that each of the subroutine runs in time
\npoly{} in the size of the current instance.
However, by Lemma~\ref{lem:running time},
the size of this instance is polynomial in $n$.
So, it is left to show that each of the subroutine is run only
polynomially many times.
Notice, that each invocation of Algorithms~\ref{pc code},
\ref{ac code}, \ref{pc code crossing} introduces a new letter to $\Sigma$,
and we know by Lemma~\ref{lem:running time} that the final size of $\Sigma$
is \polyn.
Moreover, each invocation of Algorithm~\ref{removing outer letters}
if followed by Algorithm~\ref{ac code} (in the extended version),
similarly, each Algorithm~\ref{pop code} can be associated with
Algorithm~\ref{pc code crossing} invoked after it.
And so also Algorithms~\ref{removing outer letters}, \ref{pop code}
are invoked \polyn{} times.
So the whole running time is \npoly.

It is left to show, that if the input is a DFA,
Algorithm~\ref{alg:main} can be determinised.
Firstly, notice that by
Lemmata~\ref{lem:pc noncrossing}, \ref{lem:ac inner},
\ref{lem:ai are not crossing}, \ref{lem:ac extension}, \ref{lem:pop},
\ref{lem:pc crossing}, if the instance consisted of a DFA,
each instance kept by Algorithm~\ref{alg:main} is also a DFA.
Observe, that the only non-deterministic choices in Algorithm~\ref{alg:main}
are performed when calling a subroutine
for a fully compressed membership problem for a string over an alphabet consisting
of a single letter (see Lemma~\ref{lem:unary case is in NP}).
However, the same lemma states, that when the input consists of a deterministic
automaton, the problem is in \Pclass.
Thus, there is no non-determinism in Algorithm~\ref{alg:main}.
\end{proof}

\section{Additional material for Section~\ref{sec:prelim}}

It is more convenient to represent path's list of labels as
\begin{equation*}
u_{i_1}X_{i_2}u_{i_3}X_{i_4}\cdots X_{i_{n-1}}u_{i_n},
\end{equation*}
where each $u_{i_j} \in \Sigma^*$ is a string representing the consecutive
letter labels and $X_{i_j}$ represents a nonterminal label.
Notice, that $u_{i_j}$ may be empty.
We write, that $\mathcal P$ \emph{induces} such a list of labels.
The $\eval(\mathcal P)$ \emph{defined} by such a path $\mathcal P$ is
\begin{equation*}
\eval(\mathcal P) = u_{i_1}\eval(X_{i_2})u_{i_3}\eval(X_{i_4})\cdots \eval(X_{i_{n-1}}) u_{i_n}.
\end{equation*}

\begin{proof}[Proof of Lemma~\ref{lem:unary case is in NP}]
\seepage{lem:unary case is in NP}
Notice, that if the input string $w$ is over an alphabet $\{ a \}$,
no accepting path in NFA for $w$ may use transitions that denote strings
having letters other than $a$.
Thus, any such transitions can be deleted from $N$ and we end up with
an instance, to which the result of Plandowski and Rytter%
~\cite[Theorem~5]{SLPmatchingNFA} can be applied directly.

Consider now the deterministic automaton.
As shown in the previous paragraph, we can limit ourselves to transitions
by powers of $a$.
Since the automaton is deterministic, for each state there is at most 
one transition labelled with a power of $a$,
and so the path for $w$ cycles after at most $n$ transitions.
As the length of the cycle can be calculated in \poly,
the whole problem can be easily checked in \poly.
\end{proof}

\section{Additional material for Section~\ref{sec:outline}}

\begin{proof}[Proof of Lemma~\ref{lem:different crossing}]
\seepage{lem:different crossing}
Since there are $n$ nonterminals, there are at most $n$ left-outer letters
and $n$ right-outer letters. Clearly, they can be calculated in \poly{}.

We estimate the total number of pairs
of letters that appear in $\eval(X_1)$, \ldots, $\eval(X_n)$.
Consider first such pairs that appear
in some explicit string on the right-hand side of some production.
Since the total length of explicit strings is $|G|$, there are at most $|G|$ such pairs.
Other pairs are assigned to nonterminals $X_1, \ldots , X_n$:
a pair $ab$ is assigned to $X_i$,
if it appears in $\eval(X_i)$ and it does not appear in $\eval(X_1)$, \ldots, $\eval(X_{i-1})$.
We show, that at most three different pairs are assigned to each nonterminal.
In this way, total number of different crossing-pairs is at most $|G|+3n$.
Indeed, if $ab$ is assigned to $X_i$ with a production $X_i \to u X_j v X_k$,
then $ab$ does not appear neither in $u$, $v$, as in such case it was already accounted;
nor in $\eval(X_j)$, $\eval(X_k)$, as in such case it is not assigned to $X_i$.
Thus, there are three possibilities:
\begin{itemize}
	\item $a$ is the last letter of $u$ and $b$ is the first letter of $\eval(X_j)$,
	\item $a$ is the last letter of $\eval(X_j)$ and $b$ is the first letter of $v$,
	\item $a$ is the last letter of $v$ and $b$ is the first letter of $\eval(X_k)$.
\end{itemize}
The cases, in which $u$ or $v$ is empty or there are less nonterminals on the
right-hand side of the production, are similar.

The above description can be turned to a straightforward algorithm computing
both the list of all non-crossing and crossing pairs appearing in $\eval(X_1)$, \ldots, $\eval(X_n)$.
First, the list of all pairs of letters with such appearances is calculated:
clearly, it is enough to read every rule (for $X_i$)
and store the pairs that appear in the explicit
strings and the pairs that are assigned to $X_i$.
Then, for each pair of letters it should be decided, whether it is crossing.
To this end, we check, whether it has a crossing appearance in any nonterminal
or in $N$, which can be done in \poly. Such pairs are crossing,
other are non-crossing.
%
%
%
%
\end{proof}

Instead of Lemma~\ref{polynomial non extendible appearances} we show an extended version

\begin{lemma}[stronger variant of Lemma~\ref{polynomial non extendible appearances}]
\label{polynomial non extendible appearances better}
For an inner letter $a$ and a grammar $G$, which can be given in an $a$-succinct form,
there are at most $|G|$ different lengths of $a$'s non-extendible appearances
in $\eval(X_1)$, \ldots $\eval(X_n)$.
The set of these lengths can be calculated in \poly.
\end{lemma}
\begin{proof}
\seepage{polynomial non extendible appearances}
The proof is similar to the proof of Lemma~\ref{lem:different crossing}.

Since $a$ is an inner letter, all of its non-extendible appearances
are explicit substrings in productions of $G$.
In particular, each symbol in the productions can uniquely assigned to the
non-extendible appearance to which it belongs;
this is true regardless of whether the symbol represents letter or block
of letters written in a succinct way.
Thus, there are at most $|G|$ different non-extendible appearances for $a$.
To calculate the lengths of these appearances, it is enough to read the explicit strings in the rules,
adding the appropriate lengths,
which can be done in \poly, as these lengths are at most $2^n$.
%
\end{proof}

\begin{proof}[Proof of Lemma~\ref{lem:logM iterations}]
\seepage{lem:logM iterations}
Consider any $2$ consecutive letters $ab$, where $a \neq \$$ and $b \neq \#$, appearing in the $\eval(X_n)$
at the beginning of loop starting in line~\ref{alg:mainloop}.
We show, that at least one of these two letters is compressed before the next execution
of this loop.
In this way, if we partition $\eval(X_n)$ into blocks of $4$ consecutive letters,
each block is shortened by at least one letter in each iteration of the loop from line~\ref{alg:mainloop}.
Thus the length of $\eval(X_n)$ decreases by a factor of $3/4$ in each iteration and so
this loop is executed at most $\Ocomp( n )$ times,
as in the input instance $|\eval(X_n)| \leq 2^n$.

Assume for the sake of contradiction, that none of letters $a$, $b$ is compressed during this iteration of the loop.

If $a=b$, then this pair of consecutive letters is going to be compressed,
either in line~\ref{ac noncrossing} or~\ref{ac crossing}, depending on whether $a$ is inner or outer. Contradiction.

So suppose now that $a \neq b$.
Since the pair $ab$ was not compressed before line~\ref{cr list} in Algorithm~\ref{alg:main},
it is crossing in this line and thus at least one of the letters $a$, $b$ is outer.
This outer letter is going to be compressed with the next letter
in line~\ref{pc crossing} (or earlier). Contradiction.

This ends the case inspection.

Since in the input instance it holds that $ P \leq 2^n$, this guarantees,
that the loop in line~\ref{alg:mainloop} of Algorithm~\ref{alg:main}
is run $\Ocomp(n)$ times.
\end{proof}

\section{Additional material for Section~\ref{sec:details}}

\begin{proof}[Proof of Lemma~\ref{lem:pc noncrossing}]
\seepage{lem:pc noncrossing}
The bound on the running time is obvious from the code.

Since Algorithm~\ref{pc code} only modifies the grammar by
shortening some strings in the productions (it does not create $\epsilon$-rules),
and it does not affect $\$$ and $\#$ in the rules,
the \SLPrefall{} are preserved.
The only modification in $N$ is the introduction of new transition
by a single letter (namely, by $c$) between states that are joined
by a path for $ab$.
The only change in $N$ is the introduction of a new letter transitions.
Moreover, if there is a new transition $\delta_{N'}(p',c,q')$,
then $p$ has an outgoing production by $a \notin \{\$, \#\}$, and so it was not a starting or accepting state,
and $q$ had an incoming transition by $b \notin \{\$, \#\}$, and similarly it was not a starting nor accepting state.
Thus \Autrefall{} hold for $N'$ as well.
Notice, that if $N$ is deterministic, so is $N'$:
suppose that there are two different transitions by a letter $d$ from state $p$ in $N'$.
If $d \neq c$, then these two transition are also present in $N$, which is not possible,
as $N$ is deterministic.
If $d = c$, then in $N$ there are two different paths from $p$ for a string $ab$,
which is also a contradiction.

We now show that $N'$ recognises $\eval(X_n')$
if and only if $N$ recognises $\eval(X_n)$.
To this end we demonstrate, how Algorithm~\ref{pc code} affects $\eval(X_i)$:
\begin{clm}
\label{clm:pc on compressed}
After performing Algorithm~\ref{pc code}, it holds that
\begin{equation}
\label{eq:pc on compressed}
\eval(X_i') = \PC_{ab \to c}(\eval(X_i)).
\end{equation}
\end{clm}

\begin{proof}
Notice, that as $a\neq b$, $\PC_{ab \to c}$ is well defined for each string.

The claim follows by a simple induction on the nonterminal's number:
Indeed, this is true when the production for $X_i$ has no nonterminal on the right-hand side
(recall the assumption that $a \neq b$),
as in this case the pair compression on right hand side of the production for $X_i$
is explicitly performed.
When $X_ i \to u X_j v X_k$, then
\begin{align*}
\eval(X_i)
	&=
u\eval(X_j)v\eval(X_k) \quad \text{and}\\
\eval(X_i')
	&=
\PC_{ab \to c}(u) \eval(X_j') \PC_{ab \to c}(v) \eval(X_k')\\
	&=
\PC_{ab \to c}(u) \PC_{ab \to c}(\eval(X_j')) \PC_{ab \to c}(v) \PC_{ab \to c}(\eval(X_k')),
\end{align*}
with the last equality following by the induction assumption.
Notice, that since $ab$ is a non-crossing pair, all occurrences of $ab$ in $\eval(X_i)$
are contained in $u$, $v$, $\eval(X_j)$ or $\eval(X_k)$,
as otherwise $ab$ would be a crossing pair, which contradicts the assumption.
Thus,
\begin{align*}
\PC_{ab \to c}(\eval(X_i)) = 
\PC_{ab \to c}(u) \PC_{ab \to c}(\eval(X_j')) \PC_{ab \to c}(v) \PC_{ab \to c}(\eval(X_k')),
\end{align*}
which shows that $\PC_{ab \to c}(\eval(X_i)) = \eval(X_i')$,
ending the proof of the claim.
\end{proof}

The second claim similarly establishes,
how the pair compression of a non-crossing pair affects the NFA.
To be more precise, what happens to a string defined by a path in the NFA
after applying pair compression to the underlying NFA.

\begin{clm}
\label{clm:compression goes down}
Consider a non-crossing pair $ab$
and a path $\mathcal P$ in NFA $N$, which defines a list of labels:
\begin{equation*}
u_{i_1}X_{i_2}u_{i_3}X_{i_4}\cdots X_{i_{n-1}}u_{i_n},
\end{equation*}
where each $u_{i_j} \in \Sigma^*$ is a string representing the consecutive
letter labels and $X_{i_j}$
represents a transition by a nonterminal transition.
Then
\begin{multline}
\label{eq:compression goes down}
\PC_{ab \to c}(\eval(\mathcal P)) = 
\PC_{ab \to c}(u_{i_1})\eval(X_{i_2}') \PC_{ab \to c}(u_{i_3})\eval(X_{i_4}')\cdots \eval(X_{i_{n-1}}')\PC_{ab \to c}(u_{i_n}).
\end{multline}
\end{clm}
\begin{proof}
Similarly as in Claim~\ref{clm:pc on compressed}, notice, that as $ab$ is a non-crossing pair,
the appearance of $ab$ in the string defined by $\mathcal P$ cannot
be split between a nonterminal and a string (other nonterminal).
Thus, replacement of pairs $ab$ takes place either wholly inside string $u$ or inside $\eval(X_i)$.
The former is done explicitly by $\PC_{ab \to c}$,
while~\eqref{eq:pc on compressed} establishes the form of the latter.
This ends claim's proof.
\end{proof}

Now, after proving Claims~\ref{clm:compression goes down}--\ref{clm:pc on compressed},
it is easy to show the main thesis of the lemma, i.e.
that $\eval(X_n')$ is accepted by $N'$ if and only if $\eval(X_n)$ is accepted by $N$.

\textcircled{$\Leftarrow$}
Suppose first that $\eval(X_n)$ is accepted by $N$.
Consider the accepting path $\mathcal P$ for $\eval(X_n)$, represent it as a list of labels:
$
u_{i_1}X_{i_2}u_{i_3}X_{i_4}\cdots X_{i_{n-1}}u_{i_n}
$, similarly as in Claim~\ref{clm:compression goes down}.
Of course,
\begin{equation}
\label{eq:word on a label}
\eval(\mathcal P) = \eval(X_n) = u_{i_1}\eval(X_{i_2}) u_{i_3}\eval(X_{i_4})\cdots \eval(X_{i_{n-1}})u_{i_n}.
\end{equation}
We will construct an accepting path $\mathcal P'$ in $N'$ inducing a list of labels
\begin{equation}
\label{eq:path labels pc compressed}
\PC_{ab \to c}(u_{i_1})X_{i_2}'\PC_{ab \to c}(u_{i_3})X_{i_4}'
\cdots X_{i_{n-1}}'\PC_{ab \to c}(u_{i_n}).
\end{equation}
Using~\eqref{eq:compression goes down} and recalling
that $\eval(\mathcal P) = \eval(X_n)$ will be enough to conclude that
$\eval(\mathcal P') = \PC_{ab \to c}(\eval(X_n))$.

Notice, that by Algorithm~\ref{pc code}
\begin{itemize}
	\item if there is a transition $\delta_N(p,d,q)$ for a letter $d \in \Sigma$ in $N$,
	then there is the same transition $\delta_{N'}(p,d,q)$ in $N'$.
	\item if there is a path from $p$ to $q$ for a string $ab$ in $N$
	then there is a transition $\delta_{N'}(p,c,q)$ in $N'$.
\end{itemize}
Thus, by a trivial induction on the length of the string $u$, if $\delta_{N}(p,u,q)$
then also $\delta_{N'}(p,\PC_{ab \to c}(u),q)$.
The situation is similar for nonterminals: if there is a transition $\delta_N(p,X_i,q)$ in $N$,
then there is an analogous transition $\delta_{N'}(p,X_i',q)$ in $N'$.
Thus, a path $\mathcal P'$ with the same starting and ending state as $\mathcal P$
and the list of labels as in~\eqref{eq:path labels pc compressed} is inductively defined.
Since the starting (accepting) state in $N$ and $N'$ coincide,
this shows that $\eval(\mathcal P')$ is accepted by $N'$.

\textcircled{$\Rightarrow$}
Suppose now that a string $\PC_{ab \to c}(\eval(X_n'))$ is recognised by $N'$.
Let the path of the accepting computation in $N'$ be
$\mathcal P'$, with a list of labels
\begin{equation*}
u_{i_1}'X_{i_2}'u_{i_3}'X_{i_4}'\cdots X_{i_{n-1}}'u_{i_n}'.
\end{equation*}
Similarly to the previous case, we will inductively define an accepting path $\mathcal P$
in $N$ with a list of labels
\begin{equation}
\label{eq:decompressed path}
u_{i_1}X_{i_2}u_{i_3}X_{i_4}\cdots X_{i_{n-1}}u_{i_n},
\end{equation}
where $u_{i_j}$ is obtained from $u_{i_j}'$ by replacing each $c$ by $ab$.

Notice, that by Algorithm~\ref{pc code},
\begin{itemize}
	\item if there is a letter transition $\delta_{N'}(p,c,q)$ in $N'$,
	 there is a path from $p$ to $q$ for a string $ab$ in $N$.
	\item if the a letter transition $\delta_{N'}(p,d,q)$ for a letter $d\neq c$,
	there there is the same transition $\delta_{N}(p,d,q)$ in $N$.
\end{itemize}
Thus, by a simple induction, we conclude that if there is path in $N'$ from $p$ to $q$
for a string $u_{i_j}'$, then there is a path in $N$ from $p$ to $q$ for a string $u_{i_j}$.
Now observe, that if there is a transition $\delta_{N'}(p,X_i,q)$ in $N'$,
then there is an analogous transition $\delta_{N}(p,X_i,q)$ in $N$.
This completes the construction of $\mathcal P$.
Since the starting and accepting states in $N$ and $N'$ coincide,
the constructed path $\mathcal P$ is also accepting,
furthermore, $\mathcal P$'s list of labels is as in~\eqref{eq:decompressed path}.

It is left to show, that $\eval(\mathcal P) = \eval(X_n)$.
Since $\PC_{ab\to c}$ is a one-to-one function,
it is enough to show that $\PC_{ab \to c}(\eval(\mathcal P)) = \PC_{ab \to c}(\eval(X_n))$.
Notice, that the latter equals $\eval(X_n')$, by~\eqref{eq:pc on compressed}.

Using~\eqref{eq:compression goes down} we can conclude that
\begin{align*}
\PC_{ab \to c}(\eval(\mathcal P))
	&=
\PC_{ab \to c}(u_{i_1})\eval(X_{i_2}')\PC_{ab \to c}(u_{i_3})\eval(X_{i_4}')\cdots \eval(X_{i_{n-1}})\\
	&=
u_{i_1}'\eval(X_{i_2}')u_{i_3}'\eval(X_{i_4}')\cdots \eval(X_{i_{n-1}}')u_{i_n}'\\
	&=
\eval(\mathcal P')\\
	&=
\eval(X_n'),
\end{align*}
which concludes the proof.
\end{proof}

\begin{proof}[Proof of Lemma~\ref{lem:ac inner}, stronger version]
\seepage{lem:ac inner}
We shall proof Lemma~\ref{lem:ac inner} in a little stronger version:
we allow the grammar to be in an $a$-succinct form
and NFA $N$ is assumed to satisfy the $a$-relaxed-\Autref{letter or nonterminal},
see appropriate definition between Algorithm~\ref{removing outer letters}
and Lemma~\ref{lem:ai are not crossing}.
This weaker assumption will allow proving Lemma~\ref{lem:ac extension} here.

Notice, that by Lemma~\ref{polynomial non extendible appearances better},
there are only $|G|+3n$ possible lengths of $a$'s non-extendible appearance
and that they can be calculated in \poly;
hence line~\ref{establish lengths} takes \poly.
All the loops in Algorithm~\ref{ac code}
have only polynomially many iterations.
All operations listed in Algorithm~\ref{ac code} are elementary
and can be clearly performed in \poly,
except replacing each $a^\ell$ by $a_\ell$ in a rule
in line~\ref{replace non-extendible}
and for the verification in line~\ref{transition guess},
for the former operation,
it is enough to read the rules of the grammar:
recall, that $a^{\ell}$ is represented as a pair $(a, \ell)$.
Since $\ell \leq 2^n$, addition of the lengths can be performed in \poly.
The verification is more involved,
we outline how to perform it:
For given two states $p$, $q$ and a string $a^\ell$ we want to verify,
whether there is a path from $p$ to $q$ for a string  $a^\ell$
(notice, that since $a$ is inner, none of $\eval(X_i)$ is a power of $a$).
This can be rephrased as a fully compressed membership problem for NFA
over a unary alphabet: it is enough to
\begin{itemize}
	\item restrict NFA $N$ to transitions by powers of $a$,
	\item make $p$ the unique starting state,
	\item make $q$ the unique accepting state.	
\end{itemize}
Notice, that we can restrict ourselves by transitions by powers of $a$:
since $a$ is an inner letter,
no $\eval(X_i)$ begins or ends with $a$.
Hence, when a path in the NFA defines $a^\ell$,
then this path uses only transitions by powers of $a$.

Observe, that all considered powers $a^\ell$ appear in
strings defined by $G$, and so $\ell \leq 2^n$.,
So each such $a^\ell$ can be 
represented by an SLP of polynomial size.
In particular, Lemma~\ref{lem:unary case is in NP} is applicable here
and so the verification is in \npoly.

Concerning the preservation of invariants:
we first show, that the $G'$ is not in the $a$-succinct form,
nor $N'$ is $a$-relaxed.
When Algorithm~\ref{ac code} finishes its work,
all non-extendible appearances of $a$ are replaced, in particular,
there are no succinct representations of $a$ powers inside the grammar.
This should apply to transitions labelled with powers of $a$ in $N$,
and so the last line of Algorithm~\ref{ac code} should be modified,
so that all transitions by \emph{powers} of $a$ are removed.

Now we can return to showing the preservation of the invariants:
since the only change to the productions consists of replacing
non-extendible occurrences of $a$ by a single letter,
\SLPrefall{} are preserved.
Also, the only modifications to the NFA
is the addition of new letter transitions.
Thus, \Autref{letter or nonterminal} holds.
To see that also \Autref{starting and accepting state} holds,
notice, that if $p$ receives new incoming (outgoing) transition in $N'$,
this transition is of the form $a_\ell$ and $p$
had an incoming (outgoing, respectively) transition by $a^\ell$ in $N$.
In particular, the starting and accepting state remain unaffected
and no transition by $\$$ and $\#$ are introduced.
Thus, also \Autref{starting and accepting state} holds for $N'$.

Notice, that if $N$ is deterministic, so is $N'$: suppose that there are two different
transitions by $d$ in $N'$. If $d$ is not one of the new letters, i.e. it is 
a nonterminal or an old letter, then the same transitions were present in $N$, contradiction.
If there are two transitions by a new letter $a_\ell$ from the same state $p$,
then in $N$ there are two different paths for $a^\ell$ from state $p$, contradiction.


\begin{clm}
\label{clm:ac on compressed}
After performing Algorithm~\ref{ac code}, it holds that
\begin{equation}
\label{eq:ac on compressed}
\eval(X_i') = \AC_{a}(\eval(X_i)).
\end{equation}
\end{clm}

\begin{clm}
\label{clm:appearance compression goes down}
Consider an inner letter $a$ and a path $\mathcal P$ in $N$,
which has a list of labels:
\begin{equation*}
u_{i_1}X_{i_2}u_{i_3}X_{i_4}\cdots X_{i_{n-1}}u_{i_n},
\end{equation*}
where each $u_{i_j} \in \Sigma^*$ is a string representing the consecutive
letter labels 
and $X_{i_j}$
represents a nonterminal transition,
similarly as in Claim~\ref{clm:compression goes down}.
Then
\begin{equation}
\label{eq:appearance compression goes down}
\AC_{a}(\eval(\mathcal P)) = 
\AC_{a}(u_{i_1})\eval(X_{i_2}') \AC_{a}(u_{i_3})\eval(X_{i_4}')\cdots \eval(X_{i_{n-1}}')\AC_{a}(u_{i_n}).
\end{equation}
\end{clm}

The proofs are analogous as in Lemma~\ref{lem:pc noncrossing} and are thus omitted.
Notice, that the properties stated in
Claims~\ref{clm:ac on compressed}--\ref{clm:appearance compression goes down}
do \emph{not} depend on the non-deterministic choices of Algorithm~\ref{ac code}.

It is left to show the main claims of the lemma:
$N$ recognises $\eval(X_n)$ if and only if NFA $N'$ recognises $\eval(X_n')$
for some non-deterministic choices.

\textcircled{$\Leftarrow$}
Suppose first that the $N'$ accepts $\eval(X_n')$, using the path $\mathcal P'$.
Clearly $\eval(\mathcal P') = \eval(X_n')$.
Let the list of labels on $\mathcal P'$ be
\begin{equation*}
u_{i_1}'X_{i_2}'u_{i_3}'X_{i_4}'\cdots X_{i_{n-1}}'u_{i_n}'.
\end{equation*}
Let $u_{i_j}'$ be obtained from $u_{i_j}$ be replacing each $a_{\ell_m}$ with
$a^{\ell_m}$.
We shall construct a path $\mathcal P$ in $N$, which has the same starting and ending
as $\mathcal P'$ and induces a list of labels
\begin{equation*}
u_{i_1}X_{i_2}u_{i_3}X_{i_4}\cdots X_{i_{n-1}}u_{i_n}.
\end{equation*}
Notice, that
\begin{itemize}
	\item if there is a transition $\delta_{N'}(p,X_i',q)$ in $N'$
	then there is a transition $\delta_N(p,X_i,q)$ in $N$
	\item if there is a transition $\delta_{N'}(p,b,q)$ for $b \neq a$ in $N'$,
	then there is a transition $\delta_N(p,b,q)$ in $N$
	\item if there is a transition $\delta_N(p,a_{\ell_m},q)$ in $N'$
	for some $a^{\ell_m}$ that has a non-extendible in one of $\eval(X_1)$,
	\ldots, $X_n$, then there is a path from $p$ to $q$ for a string $a^{\ell_m}$
	in $N$.
\end{itemize}
Therefore, by an easy induction, $\mathcal P$ is a valid path in $N$,
moreover, since $\mathcal P'$ is accepting, so is $\mathcal P$.
It is left to demonstrate, that $\mathcal P$ defines $\eval(X_n)$:
since $\AC_a$ is a one-to-one function,
it is enough to show that $\AC_a(\eval(\mathcal P)) = \AC_a(\eval(X_n))$.
By~\eqref{eq:ac on compressed} it holds that $\AC_a(\eval(X_n)) = \eval(X_n')$.
The value of $\AC_a(\eval(\mathcal P))$ is already known from~\eqref{eq:appearance compression goes down},
and so it is enough to show that
\begin{equation*}
\AC_{a}(u_{i_1})\eval(X_{i_2}') \AC_{a}(u_{i_3})\eval(X_{i_4}')\cdots \eval(X_{i_{n-1}}')\AC_{a}(u_{i_n}) = 
\eval(X_n').
\end{equation*}
But this is simply the fact, that path $\mathcal P'$ defines $\eval(X_n')$,
which hold by the assumption.

\textcircled{$\Rightarrow$}
Suppose now that $N$ accepts $\eval(X_n)$.
Consider the case in which Algorithm~\ref{ac code}
always made a correct non-deterministic choice,
i.e. that each time it correctly verified in line~\ref{transition guess}.

Let the accepting path $\mathcal P$ in $N$ has a list of labels
\begin{equation*}
u_{i_1}X_{i_2}u_{i_3}X_{i_4}\cdots X_{i_{n-1}}u_{i_n},
\end{equation*}
where, similarly as in Claim~\ref{clm:appearance compression goes down},
each $u_{i_j}$ is a string representing the consecutive
letter labels ($u_{i_j}$ may be empty) and $X_{i_j}$ represents a transition
by a nonterminal transition.
By the definition, 
\begin{equation*}
\eval(X_n) = u_{i_1}\eval(X_{i_2}) u_{i_3}\eval(X_{i_4})\cdots \eval(X_{i_{n-1}})u_{i_n}.
\end{equation*}
Now consider the $a$'s appearance compression applied to both side of this equality.
By~\eqref{eq:ac on compressed} and~\eqref{eq:appearance compression goes down}
\begin{equation*}
\eval(X_n') = \AC_{a}(u_{i_1})\eval(X_{i_2}') \AC_{a}(u_{i_3})\eval(X_{i_4}')\cdots \eval(X_{i_{n-1}}')\AC_{a}(u_{i_n}).
\end{equation*}
We will construct an accepting path $\mathcal P'$ with a list of labels
\begin{equation*}
\AC_{a}(u_{i_1})X_{i_2}'\AC_{a}(u_{i_3})X_{i_4}'\cdots X_{i_{n-1}}'\AC_{a}(u_{i_n}).
\end{equation*}
Notice, that $\eval(\mathcal P') = \eval(X_n')$, and so construction of
such path $\mathcal P'$ will conclude the proof.
We iteratively transform $\mathcal P$ into $\mathcal P'$.
Notice, that
\begin{itemize}
	\item if there is a transition $\delta_N(p,X_i,q)$ in $N$ then there is a transition
	$\delta_{N'}(p,X_i',q)$ in $N'$
	\item if there is a transition $\delta_N(p,b,q)$ for $b \neq a$ in $N$,
	then there is a transition $\delta_{N'}(p,b,q)$ in $N'$
	\item if there is a path in $N$ from $p$ to $q$ for string $a^{\ell}$
	that has non-extendible appearance in some $\eval(X_i)$,
	then there is a transition $\delta_{N'}(p,a_\ell,q)$ in $N'$
	(by the assumption that Algorithm~\ref{ac code} guessed correctly)
\end{itemize}
And by an easy induction $\mathcal P'$ is valid path in $N'$
and has the same starting and ending state as $\mathcal P$ .
Since $\mathcal P$ is accepting in $N$
and the starting and accepting states in $N$ and $N'$ coincide,
$\mathcal P'$ is an accepting path in $N'$.
\end{proof}

\begin{proof}[Proof of Lemma~\ref{lem:ai are not crossing}]
\seepage{lem:ai are not crossing}
We first explain, how to 
calculate the $a$-prefix ($a$-suffix); since $G$ is in $a$-succinct form,
this might be non-obvious.
It is enough to scan the explicit strings stored in the productions' right-hand sides,
summing the lengths of the consecutive $a$'s appearances.
This clearly works in \poly{} also for $G$ stored in an $a$-succinct form,
as the powers of $a$ may have length at most $2^n$, and so the length
of their representation is linear in $n$.
(the correctness of this approach is shown later).

Algorithm~\ref{removing outer letters} runs in \poly,
as the first loop has $n$ iterations and the second $|N|$ iterations and each
line can be performed in \poly.

Concerning the preservation of the invariants:
in each rule of the grammar at most $4$ non-extendible appearances of $a$
are introduced.
They may be long, however, in compressed form we treat them as singular symbols.
In this way rules of $G$ are stored in an $a$-succinct form,
which was explicitly allowed.
Then the $a$-prefix and suffix are removed and nonterminals defining $\epsilon$ are removed
from the right-hand sides of the productions.
This does not affect the \SLPrefall{}
(recall, that $a$ is not $\$$, neither $\#$).
Since the NFA is also changed, we inspect the invariants regarding $N$:
introducing new states $p_1$, $q_1$ and replacing transition
$\delta_N(p,X_i,q)$ by three transitions 
$\delta_{N'}(p,a^{\ell_i},p_1)$, $\delta_{N'}(p_1,X,q_1)$, $\delta_{N'}(q_1,a^{r_i},q)$
preserves the \Autrefall, with the exception that it $a$-relaxes
\Autref{letter or nonterminal}.

Notice, that if $N$ is deterministic, so is $N'$:
as already mentioned the only change done to $N$ is the replacement
of transition by $X_i$ by a path of three transitions,
such the the two states in the middle have exactly one incoming and outgoing transition,
which clearly preserves determinism of the automaton.

To show the correctness, we prove by induction on $i$ the two main claims of the lemma:
\begin{itemize}
	\item Algorithm~\ref{removing outer letters} correctly calculates
	the length of the $a$-prefix and $a$-suffix of $\eval(X_i)$, i.e.
	$\ell_i$ and $r_i$,
	\item $\eval(X_i) = a^{\ell_i} \eval(X_i')a^{r_i}$.
\end{itemize}
As simple corollary, observe that these two conditions imply
that $a$ is not the last, nor the first letter of $\eval(X_i')$.

For $i=1$ notice, that the whole production for $X_1$ is stored explicitly,
and so Algorithm~\ref{removing outer letters} correctly calculates the $a$-prefix
and suffix of $\eval(X_1)$ and after their removal, $\eval(X_1) = a^{\ell_1} \eval(X_1')a^{r_1}$.

For the induction step, let $X_i \to u X_j v X_k$.
Then by the induction assumption:
\begin{align*}
\eval(X_i)
	&=
u \eval(X_j) v \eval(X_k)\\
	&=
u a^{\ell_j}\eval(X_j') a^{r_j} v a^{\ell_k} \eval(X_k') a^{r_k}.
\end{align*}
There are cases to consider, depending on whether $\eval(X_j') = \epsilon$ or not
and whether $\eval(X_k') = \epsilon$ or not.
We will describe the one with $\eval(X_j') \neq \epsilon$ and $\eval(X_k') =\epsilon$,
other cases are treated in a similar way.
Then the rule is rewritten as $u a^{\ell_j}\eval(X_j') a^{r_j} v a^{\ell_k} a^{r_k}$.
Since by the inductive assumption, $a$ is not the first, nor the last letter of $\eval(X_j')$,
the $a$ prefix ($a$-suffix) of $\eval(X_j')$ is the $a$-prefix ($a$-suffix, respectively)
of $u a^{\ell_j}$ (of $ a^{r_j} v a^{\ell_k} a^{r_k}$, respectively).
Thus it is correctly calculated by Algorithm~\ref{removing outer letters}.
%
As the last action, Algorithm~\ref{removing outer letters} removes
the $a$-prefix and $a$-suffix, which shows that $a^{\ell_i}\eval(X_i')a^{r_i} = \eval(X_i)$.

It is left to show that $N$ accepts $\eval(X_n)$ if and only if $N'$
accepts $\eval(X_n')$.
To this end, notice, that the only modification to $N$
is the replacement of the transition of the form $\delta_N(p,X_i,q)$
by a path labelled with $a^{r_i},X_i',a^{\ell_i}$.
Furthermore, the two vertices inside this path have only one incoming and one outgoing
transition.
The path labelled with $a^{r_i},X_i',a^{\ell_i}$ defines the string
$a^{r_i}\eval(X_i')a^{\ell_i}$, which was already shown to be $\eval(X_i)$.
It is left to observe, that none of the newly introduced states
in the middle of the path is accepting, nor starting.
Hence the starting (accepting) states of $N$ and $N'$ coincide, 
each string is accepted by $N$ if and only if it is accepted by $N'$.
\end{proof}

\begin{proof}[Proof of Lemma~\ref{lem:ac extension}]
\seepage{lem:ac extension}
Notice, that the proof of Lemma~\ref{lem:ac inner}
was shown in a stronger version, which coincides with the statement of
Lemma~\ref{lem:ac extension}.
%
%
\end{proof}

\begin{proof}[Proof of Lemma~\ref{lem:pop}]
\seepage{lem:pop}
Majority of the proof is similar to the proof of Lemma~\ref{lem:ai are not crossing}.
Nevertheless, it is written for completeness.

The loop is executed polynomially many times,
and also each line of the code can be performed in \poly,
and so in total Algorithm~~\ref{pop code} runs in \poly.

Concerning the preservation of invariants:
since the only operation performed on $G$ is replacing nonterminal
$X_i$ by $aX_i'$ and then deleting the first letter of the nonterminal,
and nonterminals generating $\epsilon$ are explicitly removed from the rules,
the resulting grammar is in the form~\eqref{form}.
Notice, that the
invariants~\SLPref{no new nonterminals}--\SLPref{no new productions} are clearly preserved
by the listed operations.
Since the first and last symbol of $\eval(X_n)$ is not modified,
\SLPref{terminating symbols} holds as well.

Let us move to the NFA invariants: the only change applied to NFA
is the replacement of the transitions $\delta_N(p,X_i,q)$ to
by a path $\delta_{N'}(p,\first(X_i),p_1)$, $\delta_{N'}(p_1,X_i,q)$,
or by $\delta_{N'}(p,\first(X_i),q)$,
which does not affect \Autrefall.
For the same reason, if $N$ is deterministic, so is $N'$:

We first show by induction the second claim of the lemma,
i.e. that if $\eval(X_i)=bu$ for some $b \in \Sigma$, $u \in \Sigma^*$,
then $\eval(X_i')=u$.
For the induction basis consider $i=1$:
let the rule for $X_1$ be $X_1 \to bu$ for some $b\in \Sigma$ and $u \in \Sigma^*$.
Then $x_1 =b $ and $u' = u$ and the production for $X_1'$ is $X_1' \to u'$.
Clearly the claim holds in this case.

For the inductive step, consider $i>1$ and let $X_i$'s production be $X_i \to u X_j v X_k$.
Then, $\eval(X_i) = u \eval(X_j) v \eval(X_k)$ and by by induction assumption:
\begin{equation*}
u \eval(X_j) v \eval(X_k) = u \first(X_j) \eval(X_j')  v  \first(X_k)  \eval(X_k').
\end{equation*}
Since $u   \first(X_j)$ is non-empty, $\first(X_i)$ can be computed in line~\ref{first letter} of Algorithm~\ref{pop code}.
We conclude that
\begin{align*}
\eval(X_i)	
	&=
u  \first(X_j)  \eval(X_j')  v  \first(X_k)  \eval(X_k')\\
	&=
\first(X_i)  u'  \first(X_j)  \eval(X_j')  v  \first(X_k)  \eval(X_k')\\
	&=
\first(X_i)  \eval(X_i').
\end{align*}

The analysis for other forms of the rule, i.e. $X_i \to u X_j v$ or $X_i \to u$,
is similar.

We show now that $N'$ accepts exactly the same strings as $N$.
The only change done in the NFA is the replacement of transitions
of the form $\delta_N(p,X_i,q)$ by a path inducing list of labels with $\first(X_i)$
and $X_i'$ or by a transition $\delta_N(p,\first(X_i),q)$, let us consider
the former case, the latter is similar.
Notice, that $\eval(X_i) = \first(X_i)\eval(X_i')$ and so the new path
denotes the same string, as the replaced transition.
Furthermore, the newly introduced state in the middle of this path
has only one ingoing and outgoing transition.
The starting and accepting states were not modified,
and so the both automata recognise the same strings.
Since $\eval(X_n) = \eval(X_n')$, this ends the proof.

It is left to show, that none of the pairs $a_\ell b$ is crossing.
Assume for the sake of contradiction, that some of such pairs $a_\ell b$ is.
The analysis splits between cases, why the pair $a_\ell b$ is crossing.
\begin{description}
	\item[$a_\ell b$ has a crossing appearance in the NFA $N'$]
	In this case there is a state $p$ and a pair of incoming and outgoing transitions,
	such that $a_\ell$ is the last letter of the string encoded on the incoming transition
	and $b$ is the first letter on the outgoing string encoded transition.
	Moreover, at least one of this transitions is labelled with a nonterminal,
	we distinguish these two subcases:
	\begin{description}
		\item[transition into $p$ is labelled with $X_i'$]
		In such a case $a_\ell$ is the last letter of $\eval(X_i')$,
		and since $\eval(X_i) = \first(X_i) \eval(X_i')$, also the last letter of $\eval(X_i)$.
		Thus $a_\ell$ is a right-outer letter in $G$, contradiction with Lemma~\ref{lem:ac extension}.
		\item[transition into $p$ is labelled with a letter and transition from $p$ is labelled with $X_i'$]
		In this case $a_\ell$ is the label of the transition incoming to $p$.
		Observe, that due to the algorithm,
		if $X_i'$ labels the transmission from $p$,
		there is a unique transition to $p$, labelled with $\first(X_i)$.
		This means that $a_\ell$ is the first letter of $\eval(X_i)$,
		which is not possible by Lemma~\ref{lem:ac extension}.
	\end{description}
\item[$a_\ell b$ has a crossing appearance in nonterminal $X_i'$]
Let the production for $X_i'$ be $X_i' \to u X_j' v X_k'$, the cases of other productions are similar.
There are three possibilities for a crossing appearance of $a_\ell b$ in $\eval(X_i')$:
\begin{description}
	\item[$a_\ell$ is the last letter in $u$ and $b$ is the first letter in $\eval(X_j')$]
	Algorithm~\ref{pop code} replaced $X_j$ with $\first(X_j) X_j'$ in the rule for $X_i'$.
	Thus, since $a_\ell$ is the letter preceding $X_j'$ in the rule for $X_i'$,
	it holds that $a_\ell$ was the first letter in $\eval(X_j)$,
	which is a contradiction with Lemma~\ref{lem:ac extension}.
	\item[$a_\ell$ is the last letter in $\eval(X_i')$ and $b$ is the first letter in $v$]
	Since the last letter of $\eval(X_i')$ and $\eval(X_i)$ are the same,
	as $\eval(X_i) = \first(X_i)\eval(X_i')$, $a_\ell$ is the last letter of $\eval(X_i)$,
	which is a contradiction with Lemma~\ref{lem:ac extension}.
	\item[$a_\ell$ is the last letter in $v$ and $b$ is the first letter in $\eval(X_k')$]
	the proof is the same as in the first case.
	\qedhere
\end{description}
\end{description}
\end{proof}

\begin{proof}[Proof of Lemma~\ref{lem:pc crossing}]
\seepage{lem:pc crossing}
Notice first, that there are at most $|G|$ different letters $a_i$:
they were introduced by Algorithm~\ref{ac code} as representations
of non-extendible appearances of $a$, and by Lemma~\ref{polynomial non extendible appearances better}
there are at most $|G|$ different such appearances.
Other operations of Algorithm~\ref{pc code crossing} constitute of
call to Algorithm~\ref{pop code} and Algorithm~\ref{pc code},
and thus the running time and preservation of the invariants follows
properties of these two algorithms, stated in 
Lemma~\ref{lem:ac extension} and Lemma~\ref{lem:pop}
\end{proof}

\begin{proof}[Proof of Lemma~\ref{lem:running time}]
\seepage{lem:running time}
We first bound the size of $\Sigma$.
We show, that at the beginning of each
iteration of the main loop in Algorithm~\ref{alg:main}
each right-hand side of the production has at most $48n$ explicit letters,
and that inside each iteration of the main loop of Algorith~\ref{alg:main}
there are at most $12n$ new letters added
(we exclude the letter replacing compressed strings).
Notice, that during the run of Algorithm~\ref{alg:main}
grammar $G$ may be in succinct form, and accordingly we treat $a^\ell$ as one symbol.

Clearly, these bounds are true when Algorithm~\ref{alg:main} starts working.
Let us fix a rule and consider, how many new letters may be introduced in this rule.
There are only two cases, in which new letters (not coming from compression)
are introduced to a rule:
\begin{description}
	\item[changing an outer letter to an inner one (line~\ref{replace pop code} of Algorithm~\ref{removing outer letters})]
	There are at most $4$ new powers of $a$ (all possibly in succinct form)
	that may be introduced in each invocation of Algorithm~\ref{removing outer letters}.
	While these powers of $a$ are written in a succinct representation,
	they will be all replaced by single letters in the later appearance compression for $a$.
	
	In total, Algorithm~\ref{removing outer letters}
	is invoked for each outer letter, and there are at most
	$2n$ such letters, by Lemma~\ref{lem:different crossing}.
	So there are at most $8n$ new symbols introduced in this way
	for one iteration of the main loop of Algorithm~\ref{alg:main}.
	\item[popping letters (line~\ref{pop code new letter} of Algorithm~\ref{pop code})]
	Each invocation of of Algorithm~\ref{pop code} introduces
	at most $2$ new symbols. As Algorithm~\ref{pop code}
	is also used for each outer letter, in this way at most
	$4n$ new symbols were introduced per each iteration of the main loop of Algorithm~\ref{alg:main}.
\end{description}
Hence, there are at most $12n$ new letters added to the right-hand side
of a production in each iteration of the main loop in Algorithm~\ref{alg:main}.
Still, the main task performed by Algorithm~\ref{alg:main} is the compression:
an argument similar as in the proof Lemma~\ref{lem:logM iterations}
can be used to show that the size of the explicit strings in the rules
decreases by a factor of $3/4$ in each iteration of loop from line~\ref{alg:mainloop}
in Algorithm~\ref{alg:main}.
Of course, the newly introduced letters may be unaffected by this compression.
It is left to verify, that $48n$ is indeed the upper bound on the size of the right-hand size
of a rule, let it be $X_i \to \alpha_i$:
\begin{equation*}
|\alpha_i'| \leq \frac{3}{4} \cdot |\alpha_i| + 12n \leq \frac{3}{4} \cdot 48n + 12n \leq 48n.
\end{equation*}
Which proves the bound on $|G|$ at the end of each iteration of the main loop.
Notice, that as there at most $12n$ new letters added to this rule, inside the phase
the size of $G$ is at most $60n^2$.

We now turn our attention to the size of $\Sigma$.
Again, consider the execution of Algorithm~\ref{alg:main} and one iteration of the main loop.
We show, that there are \polyn{} letters introduced in one such iteration.
New letters are introduced when compression of pairs or appearance compression is applied.
There are the following possibilities
\begin{description}
	\item[compression of an inner letter (line~\ref{pc new letter} of Algorithm~\ref{pc code})]
	Each compression of an inner letter
	decreases the total length of explicit strings used in $G$
	by at least $1$. Since the size of each right-hand side is at most $48n$ at the beginning
	of the iteration and there are at most $12n$ new letters introduced to a rule in each iteration,
	there can be at most $60n^2 $ such compressions.
	\item[compression of a non-crossing pair (line~\ref{replace non-extendible} of Algorithm~\ref{ac code})]
	The same argument as in the previous case applies.
	\item[appearance compression for of an outer letter (Algorithm~\ref{removing outer letters} followed by Algorithm~\ref{ac code})]
	There are at most $2n$ outer letters,
	by Lemma~\ref{lem:different crossing}, and each of them has
	at most $|G|$ different lengths of non-extendible appearances,
	by Lemma~\ref{polynomial non extendible appearances better}.
	Compressing all of them introduces at most $2n|G|$ new letters to $\Sigma$.
	\item[compression of a crossing pair (call to Algorithm~\ref{pc code} made by Algorithm~\ref{pc code crossing})]	
	The crossing pairs compression is run for each of the $2n$ outer letters.
	When such an outer letter is fixed,
	there are at most $|G|+3n$ different pairs of letters
	appearing in $\eval(X_n)$, by Lemma~\ref{lem:different crossing},
	this may introduce up to $|G|+3n$ new letters.
	Hence, in total this might introduce up $2n^|G|+6n^2$ new letters to $\Sigma$.
\end{description}
Thus, the size of $|\Sigma|$ is \polyn.

It is left to bound the size of $|N|$:
as there are at most $n$ non-terminal transitions,
the the size of transition function of $N$ is at most
$\Ocomp(|Q_N|^2 \cdot (|\Sigma| + n))$.
So it enough to bound the number of states of $N$ by \polyn.
Recall, that for simplicity we assumed that all transitions
in the input DFA are labelled with different SLPs, in particular,
this polynomial bound holds for the input instance.

There are only two situations, in which new states are introduced:
\begin{description}
	\item[changing $a$ from outer letter to inner one (Algorithm~\ref{removing outer letters} in line~\ref{new states outer letters})]
	Algorithm~\ref{removing outer letters} introduces two states
	per nonterminal transition,
	and there are at most $n$ such transitions, by~\Autref{letter or nonterminal}.
	So it is enough to estimate, how many times Algorithm~\ref{removing outer letters}
	is invoked.
	Algorithm~\ref{removing outer letters} is run for each outer letter,
	and there are at most $2n$ outer letters.
	Thus, in one iteration of the main loop of Algorithm~\ref{alg:main}
	it adds at most $\Ocomp(n^2)$ states in total in this case.
	\item[popping letters (Algorithm~\ref{pop code} in line~\ref{new states pop})]
	A similar argument applies, with the only difference,
	that Algorithm~\ref{pop code} introduces one new state per nonterminal transition.
\end{description}
It is left to recall, that by Lemma~\ref{lem:logM iterations}
the main loop of Algorithm~\ref{alg:main} is run
$\Ocomp(n)$ times.
This ends the proof of the lemma.
\end{proof}

\end{document}